\documentclass{ws-mpla}

\begin{document}

\renewcommand{\draftnote}{}
\renewcommand{\trimmarks}{}

\markboth{E. A. Matute} {Neutrino mass generation from the
perspective of presymmetry}

\catchline{}{}{}{}{}

\title{Neutrino mass generation from the
perspective of presymmetry}

\author{\footnotesize Ernesto A. Matute}

\address{Departamento de F\'{\i}sica, Universidad de Santiago de
Chile,\\ Usach, Casilla 307, Santiago, Chile\\
ernesto.matute@usach.cl}

\maketitle

\pub{}{}

\begin{abstract}
The Standard Model (SM) with one right-handed neutrino per
generation is revisited with presymmetry being the global
$U(1)_{B-L}$ symmetry of an electroweak theory of leptons and
quarks with initially postulated symmetric fractional charges. The
cancellation of gauge anomalies and the non-perturbative
normalization of lepton charges proceed through the mixing of
local and topological charges, the global $B-L$ measuring the
induced charge associated with a unit of topological charge, and
the mathematical replacement of the original fractional charges
with the experimentally observed ones. The $U(1)_{B-L}$ symmetry
of the SM with Dirac neutrinos is seen as a residual presymmetry.
High-scale and low-scale seesaw mechanisms proposed to explain the
mass of neutrinos are examined from the perspective of
presymmetry, be they of Majorana or pseudo-Dirac type. We find
that the tiny mass splitting in pseudo-Dirac neutrinos and the
mass of heavy neutrinos ride on the opposite ends of the seesaw.
We show that pseudo-Dirac neutrinos contain extra sterile
neutrinos with imprints of presymmetry and for heavy ones we get
constraints favoring the low-scale linear seesaw over the inverse
variant.

\end{abstract}

\ccode{PACS Nos.: 14.60.St, 14.60.Pq, 11.30.Hv, 11.30.Ly}

\section{Introduction}
\label{introduction}

Neutrino oscillations, which can be understood if neutrinos have
tiny masses, are well established experimental evidences for
physics beyond the Standard Model (SM).\cite{PDG} The SM predicts
massless neutrinos and conservation of the global baryon minus
lepton ($B-L$) quantum number because it includes exclusively
renormalizable terms, only a Higgs doublet, and no right-handed
(RH) sterile neutrino. The most natural and straightforward
approach to generate neutrino masses and mixing is then to extend
the SM with RH neutrinos having Yukawa couplings just as in the
case of charged leptons and quarks. This extension, however, does
not explain the smallness of neutrino masses in comparison with
all other massive fermions. Yet, the lepton--quark symmetry
exhibited plainly in the electroweak sector of the SM, when {\it
three} RH neutrinos are added, together with the conservation of
$B-L$, are strong motivations for
presymmetry.\cite{Presym1,Presym2}

The goal of this paper is to revise the conception of presymmetry
and show its relevance in the mechanisms of neutrino mass
generation, giving clues to discriminate scenarios proposed in the
literature where the nature and flavor structure of neutrinos is
predicated. Here it is described as the global $U(1)_{B-L}$
symmetry of an electroweak theory of leptons and quarks with
initially postulated symmetric fractional charges, where one RH
neutrino per generation is considered. The cancellation of gauge
anomalies and the non-perturbative normalization of lepton charges
proceed via the mixing between local and topological charges, the
global $B-L$ being attached to the induced charge associated with
a unit of topological charge, and the mathematical replacement of
the original fractional charges with the experimentally observed
ones. This new implementation of presymmetry, which modifies the
earliest form of the model proposed in Ref.~\refcite{Presym2},
allow us to address the problem of neutrino mass generation in a
more conventional way. Thus, we regard the global $U(1)_{B-L}$
symmetry of the SM extended with Dirac neutrinos as a residual
presymmetry, even though their small masses remain unexplained.

One of the simplest ways to create a strong mass hierarchy between
neutrinos and charged leptons is to use the seesaw mechanism at a
high scale with the Majorana mass terms of RH
neutrinos,\cite{Seesaw1}$^{\mbox{--}}$\cite{Seesaw5} breaking the
residual $B-L$ symmetry of the SM with Dirac neutrinos and
producing light and heavy Majorana neutrinos. Nevertheless, there
is no positive experimental signatures for them, so that light or
heavy pseudo-Dirac neutrinos may exist. Popular models are the
low-scale inverse\cite{Inverse1}$^{\mbox{--}}$\cite{Inverse3} and
linear\cite{Linear1}$^{\mbox{--}}$\cite{Linear3} altered versions,
in which a second set of sterile neutrinos is added to the
particle content that mixes almost maximally with the first set of
gauge singlet neutrinos to form heavy pseudo-Dirac neutrinos, and
generate light Majorana neutrinos.

We use presymmetry to inquire into the mass terms of seesaw
mechanisms for light neutrinos characterized by fermionic singlet
extensions within the SM gauge group. The new restriction is the
absence of the Majorana mass of the extra sterile neutrinos, which
discards the low-scale inverse seesaw in favor of its linear
variant. The inverted high-scale seesaw mechanism generating light
pseudo-Dirac neutrinos and heavy Majorana neutrinos is also
considered and treated in a similar way, even though there is some
tension with cosmological bound on the number of the relativistic
neutrino species.\cite{Neff} We take it as an example to show how
constraints from presymmetry translate into constraints on the
generation of light neutrino masses. We find that these
pseudo-Dirac particles contain the adulterant, extra sterile
neutrinos with the distinguishing features of presymmetry. A
characteristic of these seesaws is that the tiny mass splitting in
pseudo-Dirac neutrinos and the mass of heavy neutrinos ride on the
opposite ends. Thus, any phenomenological bound on the mass
splitting necessarily implies an adjustment of the seesaw scale.
However, the pseudo-Dirac character of the neutrino mass may be
difficult to probe if its mass splitting comes out to be very
tiny. Remember that the zero limit of such a splitting leads to
Dirac neutrinos.

The presymmetry proposed in this paper is mostly used for the
sector of extra sterile neutrinos, working in the framework of the
seesaw mechanism which allows us to understand the observed
smallness of active neutrino masses. However, this is not
sufficient to explain their observed mixing pattern, so that
further input is required. A possible way of finding a solution to
this problem is to extend the SM symmetry with a discrete flavor
symmetry.\cite{King} This may be supplemented with an extended
gauge symmetry, including a gauged presymmetry, i.e. promoting the
global $U(1)_{B-L}$ of presymmetry to a local symmetry. This
issue, which has been discussed extensively in the literature, is
beyond the scope of this work and we will not elaborate on it any
further.

We organize the presentation of this paper as follows. In Sec.~2,
we revise presymmetry in the SM extended with three RH neutrinos,
invoking the global $U(1)_{B-L}$ symmetry, connecting $B-L$ to the
induced charge associated with topological charge, and normalizing
non-perturbatively charges in leptons to integer values through
the mathematical replacement of the initially postulated
fractional charges with the experimentally observed charges. In
Sec.~3, we look on the role of presymmetry in the low-scale linear
seesaw mechanism without considering gauge symmetries beyond the
SM, leading to light Majorana neutrinos and adulterated heavy
pseudo-Dirac neutrinos as the active left-handed (LH) neutrinos
are substituted by the extra ones with the same chirality. In
Sec.~4, we focus on its inverted high-scale alternative producing
adulterated light pseudo-Dirac neutrinos with extra sterile RH
neutrinos instead of the ordinary RH ones. Regarding the notation
used in the text, we mention that the extension of the SM with
sterile fermions which play the role of the ordinary RH neutrinos
and the extra LH or RH neutrinos destined to adulterate the
pseudo-Dirac neutrinos, has motivated us to denote them as
$\nu_R$, $\nu_L^\prime$, and $\nu_R^\prime$, respectively.
Constraints on neutrino masses in these seesaw mechanisms are
given in Sec.~5, based on the experimental bounds established for
the mass splitting in pseudo-Dirac neutrinos. The conclusions are
summarized in Sec.~6.
%%%%%%%%%%%%%%%%%%%%%%%%%%%%%%%%%%%%%%%%%%%%%%
\section{Presymmetry with Three Right-Handed
Neutrinos} \label{presymmetry}

We partially review presymmetry in order to implement the changes
in its definition. The key indications that make presymmetry
apparent are the existence of symmetries between the standard
hypercharges of chiral leptons and quarks when one RH neutrino is
added in each of the three fermionic families of the SM, which
can be expressed as
\begin{eqnarray}
\begin{array}{ll}
Y(\nu_{L}) = Y(u_L) + \Delta Y(u_L) = -1 \, , \qquad & Y(\nu_{R})
= Y(u_R) + \Delta Y(u_R) = 0 \, , \\ [10pt] Y(e_L) = Y(d_L) +
\Delta Y(d_L) = -1 \, , \qquad & Y(e_R) = Y(d_R) + \Delta Y(d_R) =
-2 \, ,
\end{array}
\label{Ysym1}
\end{eqnarray}
where the $\Delta Y=-4/3$ for quarks is a \emph{global} fractional
piece of hypercharge connected to lepton and baryon numbers by
\begin{equation}
\Delta Y(q) = - 4 \; (B-L)(q) \, . \label{DeltaY}
\end{equation}
They can also be written as
\begin{eqnarray}
\begin{array}{ll}
\displaystyle Y(u_L) = Y(\nu_{L}) + \Delta Y(\nu_{L}) =
\frac{1}{3} \, , \qquad & \displaystyle Y(u_R) = Y(\nu_{R}) +
\Delta Y(\nu_{R}) = \frac{4}{3} \, , \\ [10pt] \displaystyle
Y(d_L) = Y(e_L) + \Delta Y(e_L) = \frac{1}{3} \, , \qquad &
\displaystyle Y(d_R) = Y(e_R) + \Delta Y(e_R) = - \frac{2}{3} \, ,
\end{array}
\label{Ysym2}
\end{eqnarray}
with $\Delta Y=4/3$ for leptons, where, however,
Eq.~(\ref{DeltaY}) must have $L/3$ instead of $L$. Here the
conventional relation $Q=T_3+Y/2$ between electric charge, weak
isospin, and hypercharge, is used. Any other hypercharge
normalization can change the value of the global part $\Delta Y$,
but the charge symmetry described in Eqs.~(\ref{Ysym1}) and
(\ref{Ysym2}) is maintained.\cite{Presym3}

Presymmetry has to do with the equality of lepton and quark
charges if the global part $\Delta Y$ is kept out of sight. The
inclusion of one RH neutrino per generation is indispensable to
completing this correspondence between charges and a symmetry of
lepton and quark contents.

The charge symmetry reflected in Eqs.~(\ref{Ysym1}) and
(\ref{Ysym2}) is not accidental. These equations can be understood
as manifestations of a charge normalization involving the
so-called preleptons and prequarks.\cite{Presym2} These are
defined by the quantum numbers of the respective leptons and
quarks, excluding charge values. Preleptons and prequarks have the
same hypercharges as their quark and lepton weak partners,
respectively. We remark, however, that Eqs.~(\ref{Ysym1}) and
(\ref{Ysym2}) do not mean physical charge fractionalization in
lepton or quarks, and that preleptons and prequarks are not
physical dynamical entities. They are simply the initial lepton
and quark states considered as mathematical entities,
unrenormalized charged elements of the field theory, out of which
physical particles are built up.

The question now is, what prearrangement should be chosen, the
prelepton--quark or the lepton--prequark? It was not asked before in
Ref.~\refcite{Presym2}. As we shall see, the seesaw mechanism to
explaining the smallness of neutrino mass calls for the
prelepton--quark scheme. So, interestingly enough, fractional
charges are ``hidden'' not only in hadrons, but also in leptons.
This gives certain symmetry to particles that occur in nature,
though the ``hidden'' fractional charges in leptons established by
presymmetry are unphysical. Otherwise, it would contradict
precision electroweak measurements as well as several measurements
at the Large Hadron Collider (LHC).

Consequently, using a hat accent over the corresponding flavor
symbol to denote preleptons, we rewrite Eq.~(\ref{Ysym1}) as
\begin{eqnarray}
\begin{array}{l}
Y(\check{\nu}_{L,R}) = Y(\hat{\nu}_{L,R}) + \Delta
Y(\hat{\nu}_{L,R}) \, , \\ [10pt] Y(\check{e}_{L,R}) =
Y(\hat{e}_{L,R}) + \Delta Y(\hat{e}_{L,R}) \, ,
\end{array}
\label{Ysym3}
\end{eqnarray}
with prelepton--quark charge symmetry described as
\begin{eqnarray}
\begin{array}{ll}
Y(\hat{\nu}_{L,R}) = Y(u_{L,R}) , \qquad & \Delta
Y(\hat{\nu}_{L,R}) = \Delta Y(u_{L,R}) , \\ [10pt]
Y(\hat{e}_{L,R}) = Y(d_{L,R}) , \qquad & \Delta Y(\hat{e}_{L,R}) =
\Delta Y(d_{L,R}) ,
\end{array}
\label{lqPresym1}
\end{eqnarray}
where $\check{\nu}_{L,R}$, $\check{e}_{L,R}$, with a check accent,
refer to preleptons with normalized integer charges, which in the
end will be mathematically replaced by the standard leptons
$\nu_{L,R}$, $e_{L,R}$, respectively. The connection of $\Delta Y$
for preleptons with the lepton and baryon numbers in these cases is
\begin{equation}
\Delta Y(\hat{\ell}) = - 4 \; (B-L)(\hat{\ell}) \, ,
\label{Delta2}
\end{equation}
with $\Delta Y=-4/3$ and $L=-1/3$, the $3$ being attributed to the
number of families.\cite{Presym2} In this way, Eq.~(\ref{DeltaY})
applies to quarks as well as preleptons. And as exposed below,
Eq.~(\ref{Ysym3}) reflects a non-perturbative charge normalization
having two fractional parts, associated with local and topological
properties of fields.\cite{Wilczek} At this point, we remark that
the appearance of the global charge-shift in Eq.~(\ref{Delta2}),
which normalizes charges to the SM values, modifies the form
presented in Ref.~\refcite{Presym2}; Eqs.~(\ref{DeltaY}) and
(\ref{Delta2}) do not apply to leptons and prequarks. If $B-L$ is
the meaningful quantum number to consider in $\Delta Y$, we find
here another reason to choose the prelepton--quark instead of the
lepton--prequark framework. Besides, as described in the
following, $B-L$ is associated with the global $U(1)$ of
presymmetry.

The interactions of preleptons and quarks with the gauge fields
are assumed to be described by the same Lagrangian of the gauge
sector of the SM with leptons and quarks excepting hypercharge
couplings. In the scenario of preleptons, Majorana mass terms are
forbidden for active RH preneutrinos because these have nonzero
hypercharge, but they are a possibility at the physical
lepton--quark level, where they become sterile. More specifically,
restricting ourselves to the case of one generation, the Yukawa
Lagrangian with Majorana mass for the RH neutrino is given by
\begin{equation}
- {\cal L}_{\nu} = y \, \bar{\ell}_L \tilde{\phi} \nu_R +
\frac{1}{2} \, m_R \, \bar{\nu}^c_L \nu_R + h.c. \, ,
\label{Yukawa}
\end{equation}
where $\ell_L=(\nu_L,e_L)^T$ is the lepton doublet,
$\tilde{\phi}=i \sigma_2 \phi^\dagger$, $\phi=(\phi^+,\phi^0)^T$
is the Higgs doublet, $m_R$ refers to the Majorana mass term, and
$\nu^c_R=C\bar{\nu}^T_L$.\cite{Langacker} The Yukawa coupling for
preneutrinos is instead
\begin{equation}
- {\cal L}_{\hat{\nu}} = \hat{y} \, \bar{\hat{\ell}}_L
\tilde{\phi} \hat{\nu}_R + h.c. \, . \label{preyukawa}
\end{equation}
Since the Majorana mass terms are required to vanish at this
underlying level of prefermions, the model is invariant under the
global $U(1)_{B-L}$ symmetry, with preleptons (and quarks) in
doublets and singlets transforming as $B-L=1/3$, the $3$ related
to the number of fermionic families.\cite{Presym2} Presymmetry
then includes this symmetry with charge assignments established by
the $B-L$ charge symmetry between preleptons and quarks. It is
worth emphasizing that this scheme demands only one RH neutrino
per generation.

Thus, in this new implementation, presymmetry is defined by the
invariance of the initial electroweak Lagrangian under the global
$U(1)_{B-L}$ symmetry with preleptons and quarks having $B-L=1/3$.
The charge shift $\Delta(B-L)=-4/3$ for preleptons, breaking the
charge symmetry between preleptons and quarks, leads to the
standard charges observed in leptons.

We now explain the origin of the induced charge given in
Eq.~(\ref{Delta2}). It is related to the fact that the nonstandard
hypercharges of the original fermionic states lead to gauge
anomalies in the couplings by fermion triangle loops of three
currents related to the chiral $\mbox{U(1)}_Y$ and
$\mbox{SU(2)}_L$ gauge symmetries. Following
Ref.~\refcite{Presym2} in the framework of preleptons and quarks,
the $\mbox{U(1)}_Y$ gauge current in all representations
\begin{equation}
\hat{J}^{\mu}_{Y} = \overline{\hat{\ell}}_{L} \gamma^{\mu}
\frac{Y}{2} \hat{\ell}_{L} + \overline{\hat{\ell}}_{R}
\gamma^{\mu} \frac{Y}{2} \hat{\ell}_{R} + \overline{q}_{L}
\gamma^{\mu} \frac{Y}{2} q_{L} + \overline{q}_{R} \gamma^{\mu}
\frac{Y}{2} q_{R} \, , \label{gaugecurrent}
\end{equation}
exhibits the $\mbox{U(1)}_Y[\mbox{SU(2)}_L]^2$ and
$[\mbox{U(1)}_Y]^3$ anomalies due to the nonvanishing of the
following sums which include one RH preneutrino in each
generation:
\begin{equation}
\sum_{L} Y = 8 \, , \qquad \sum_{LR} Y^{3} = - 24 \, ,
\label{sums}
\end{equation}
where the first runs over the LH and the second over the LH and RH
topological preleptons and quarks, with $(-1)$ for the RH
contributions. Their cancellations can be done with a counterterm
which contains topological currents or Chern--Simons classes
associated with the $\mbox{U(1)}_Y$ and $\mbox{SU(2)}_L$ gauge
groups:
\begin{equation}
J^{\mu}_{T} = \frac{1}{4} \, K^{\mu} \sum_{L} Y + \frac{1}{16} \,
L^{\mu} \, \sum_{LR} Y^{3} \, , \label{currX}
\end{equation}
where
\begin{eqnarray}
\begin{array}{l}
\displaystyle K^{\mu} = \displaystyle \frac{g^{2}}{8 \pi^{2}} \,
\epsilon^{\mu\nu\lambda\rho} \; \mbox{tr} \!\! \left( W_{\nu}
\partial_{\lambda} W_{\rho} - \frac{2}{3} \, i g
W_{\nu} W_{\lambda} W_{\rho} \right) , \\ [10pt] \displaystyle
L^{\mu} = \displaystyle \frac{{g'}^{2}}{12 \pi^{2}} \,
\epsilon^{\mu\nu\lambda\rho} A_{\nu}
\partial_{\lambda} A_{\rho} \, ,
\end{array}
\label{CS}
\end{eqnarray}
so that the new current $J^{\mu}_{Y} = \hat{J}^{\mu}_{Y} +
J^{\mu}_{T}$ is anomaly-free, gauge noninvariant, and symmetric
under the exchange of preleptons and quarks. Furthermore, its
charge is not conserved because of the topological charge
associated with a weak instanton, which is related in turn to the
change in the topological winding number of the asymptotic, pure
gauge field configurations, assuming that the spacetime region of
nonzero energy density is bounded. In fact, endorsing the
principle of equality for all preleptons of the system in the
partition of the topological charge, the change in each charge,
using Eqs.~(\ref{sums})--(\ref{CS}) for the pure gauge fields,
is\cite{Presym2}
\begin{equation}
\Delta Y = \displaystyle \frac{1}{3} \, n \, , \label{deltaQ}
\end{equation}
where the topological charge is defined by
\begin{equation}
n = \int d^{4}x \, \partial_{\mu} K^{\mu} = \frac{g^{2}}{16
\pi^{2}} \int d^{4}x \, \mbox{tr} (W_{\mu\nu} \tilde{W}^{\mu\nu})
\, , \label{QT1}
\end{equation}
$W_{\mu\nu}$ denoting the $SU(2)_L$ field strength. This
topological index becomes zero in the $\mbox{U(1)}_Y$ case.

Vacuum states of different topological numbers are therefore
tunneled by $\mbox{SU(2)}_L$ instantons bearing topological
charges, making possible in principle the charge shifts and
transitions from fermions with nonstandard charges to those with
standard charges. Each hypercharge is shifted by the same amount,
which can be written as
\begin{equation}
\Delta Y = n \, (B-L) \, . \label{norma}
\end{equation}

We then have a situation where there are two charges whose mixing
defines conventionally normalized charges according to
\begin{equation}
Y_{SM} = Y + n \, (B-L) \, ,
\end{equation}
the first being local and the second topological. The attached
$B-L$ in a sense measures the induced charge associated with a
unit of topological charge.

Quarks, being topologically trivial, have $n=0$, while in the case
of preleptons the topological $n=-4$ configuration yielding
Eq.~(\ref{Delta2}) is determined by the cancellation of anomalies
and removal of the associated counterterm (see Eq.~(\ref{currX})).
Here we remark that the counterterm is used to renormalize the
charges and remove the gauge anomalies, without introducing new
fermions having suitable quantum numbers under the gauge groups as
usually done. The renormalized fields and anomaly-free new
fermions are the physical leptons themselves.

Preleptons have a vacuum gauge field configuration of winding
number $n_{-}=4$, if gauge freedom is used to set $n_{+}=0$ for
leptons. The transformation of preleptons into leptons is by means
of an Euclidean topological weak instanton with topological charge
$n=-4$, which in Minkowski spacetime is regarded as a quantum
mechanical tunneling event between vacuum states of weak
$\mbox{SU(2)}_L$ gauge fields with different topological winding
numbers. In this manner, preleptons and leptons are differentiated
by the topological vacua of their weak gauge configurations,
tunneled by a weak four-instanton carrying the topological charge
and inducing the global fractional piece of charge needed for
normalization. We then represent the hypercharge fractionalization
in preleptons described in Eq.~(\ref{Ysym3}) as follows:
\begin{eqnarray}
\begin{array}{ll}
\check{\nu}^{-1}_L = \left( \hat{\nu}^{+\frac{1}{3}}_L
\right)^{-\frac{4}{3}} , \qquad & \check{\nu}^{0}_R = \left(
\hat{\nu}^{+\frac{4}{3}}_R \right)^{-\frac{4}{3}} , \\ [10pt]
\check{e}^{-1}_L = \left( \hat{e}^{+\frac{1}{3}}_L
\right)^{-\frac{4}{3}} , \qquad & \check{e}^{-2}_R = \left(
\hat{e}^{-\frac{2}{3}}_R \right)^{-\frac{4}{3}} ,
\end{array}
\end{eqnarray}
exhibiting the preleptons with fractional charge, the global
induced hypercharge $\Delta Y = - 4 / 3$ associated with the
topological charge, and the hypercharge symmetry between
preleptons and quarks as given in Eq.~(\ref{Ysym2}). Again, this
representation cannot be taken as an actual fragmentation of
charge in leptons.

Within the framework of the initial lepton--quark charge symmetry
with one RH neutrino per generation, the global piece of
hypercharge has a weak topological character. It has been pointed
out, however, that any weak topological property cannot have
observable effects at the zero-temperature scale due to the
smallness of the weak coupling. This reaffirms the idea that the
charge structure in Eq.~(\ref{Ysym3}) simply does reflect a charge
non-perturbative normalization involving states with topological
features.

Thus, the transitions from preleptons to leptons via the weak
$\mbox{SU(2)}_L$ instantons do not take place in the actual world
because preleptons are not physical dynamical entities, i.e. they
cannot be discovered experimentally; they are simply the initial
states of the theory whose fractional charge has to be normalized
to integer values non-perturbatively. In a sense, such
transformations are frustrated by the extreme smallness of the
instanton transition probability at zero temperature, and the
charge normalization removes the extraordinarily large time scale
for them, allowing for preleptons with trivial topology and
standard charges, which mathematically become the leptons to begin
with in the usual local quantum field study, at the next effective
level of less complexity description. Quantum fields anomalies are
absent to all order of perturbation theory once the charge
normalization is realized.

Yet, there is a proper symmetry transformation at the level of
preleptons defined by presymmetry between preleptons and quarks in
the electroweak sector of the Lagrangian, which demands a
correspondence between fermionic contents at the stages of
preleptons and leptons. We then get a picture where colorless
preleptons with fractional electric charge are ``hidden'' because
of their nontrivial topology, while topologically trivial quarks
also with fractional charge are hidden because of their color
charge. They build up integer charged, topologically trivial and
colorless particles.

On the other hand, at the topologically trivial lepton level with
particles being produced by local action of fields, where the
charge symmetry $B-L$ and so presymmetry are broken, the inclusion
of Majorana mass terms for RH neutrinos violating the associated
$U(1)_{B-L}$ symmetry as in Eq.~(\ref{Yukawa}), is a plain
possibility. In principle, in a bottom-up approach, the coupling
constant in these terms, which is independent of active neutrinos,
can have any value. As a matter of fact, small Majorana mass terms
are used in the pseudo-Dirac regime, while large ones are
considered in the seesaw limit. But, the pseudo-Dirac option is
objected because it does not explain the tiny mass of neutrinos
and the high-scale seesaw is questioned since it cannot be tested.
No residual signatures of presymmetry appear in either of these
models.
%%%%%%%%%%%%%%%%%%%%%%%%%%%%%%%%%%%%%%%%%%%%%%
\section{Presymmetry in Low-Scale Seesaw with Heavy \\
Pseudo-Dirac Neutrinos} \label{linearseesaw}

Presymmetry in the scenario of preleptons and quarks
with symmetric fractional electroweak charges gives room to manage
the seesaw mechanism at a low scale. We note that in case the
framework of prequarks and leptons be chosen, presymmetry would
imply that neutrinos are Dirac fermions without explaining their
small mass, since only prequarks would be subject to charge
normalization and Majorana mass terms would be forbidden by the
$U(1)_{B-L}$ symmetry.

The extension of the SM by means of sterile neutrinos with Dirac
and Majorana mass terms allows to have masses in a generic form.
The scenario to be adopted here plays a heavy pseudo-Dirac
neutrino as in the usual low-scale seesaws after spontaneous
symmetry breaking.\cite{Inverse1}$^{\mbox{--}}$\cite{Linear3} In
order to realize these seesaws, the SM is extended by adding one
RH ($\nu_R$) and one LH ($\nu^\prime_L$) gauge singlet neutrino in
each generation of leptons and quarks. Restraining ourselves for
simplicity and purposes of this paper to the instance of one
generation (it is straight forward to extend the results to three
generations), the Yukawa Lagrangian for the neutrino sector of the
SM with the two sterile neutrinos is
\begin{equation}
- {\cal L}_{\nu} = y \, \bar{\ell}_L \tilde{\phi} \nu_R + y^\prime
\, \bar{\ell}_L \tilde{\phi} \nu^{\prime c}_R + \frac{1}{2} \left(
\begin{array}{cc} \bar{\nu}^c_L & \bar{\nu}^\prime_L
\end{array} \right) \left(
\begin{array}{cc}
m_R & \mu^\prime_D \\
\mu^\prime_D & m^\prime_R
\end{array} \right)
\left( \begin{array}{c} \nu_R \\ \nu^{\prime c}_R
\end{array} \right)
+ \; h.c. \, , \label{YukawaL2}
\end{equation}
where the Majorana mass terms with $m_R$ and $m_R^\prime$, and the
Dirac mass $\mu^\prime_D$ between $\nu^\prime_L$ and $\nu_R$, are
included. Although they can be generated from the vacuum
expectation value of gauge singlet scalars that couple to the
sterile neutrinos, in this work we just follow the
phenomenological choice defined by the effective Lagrangian in
Eq.~(\ref{YukawaL2}). The extra Majorana mass term
$\bar{\nu}^{\prime c}_R \nu^\prime_L$, allowed by gauge
invariance, is not permitted by presymmetry. In fact, the
Lagrangian underlying Eq.~(\ref{YukawaL2}) at the prelepton level
is
\begin{equation}
- {\cal L}^{\prime}_{\hat{\nu}} = \hat{y} \, \bar{\hat{\ell}}_L
\tilde{\phi} \hat{\nu}_R  + h.c. \, , \label{preyukawa2}
\end{equation}
which is identical to Eq.~(\ref{preyukawa}). In this Lagrangian,
the Yukawa term $\bar{\hat{\ell}}_L \tilde{\phi} \nu^{\prime c}_R$
and the mass term $\bar{\nu}^\prime_L \hat{\nu}_R$, involving the
extra sterile neutrino of trivial topology, are not admitted by
gauge invariance. The extra Majorana mass term $\bar{\nu}^\prime_L
\nu^{\prime c}_R$, consistent with gauge invariance, is not
permitted by the $U(1)_{B-L}$ of presymmetry. Moreover, this term
does not include presymmetric fields subject to charge
normalization, so that, advocating consistency, it also has to be
eliminated from Eq.~(\ref{YukawaL2}), enforcing $m^\prime_R =0$.
Besides, a symmetry of particle content between preleptons and
quarks appears in the Lagrangian, since the extra sterile neutrino
has been removed by presymmetry.

The spontaneous symmetry breaking of the electroweak gauge group
of the SM through the vacuum expectation value of the neutral
component of the Higgs doublet generates the Dirac mass
$m_D=y\langle \phi^0 \rangle$ and the mixing coupling
$m^\prime=y^\prime\langle \phi^0 \rangle$, which is not a Dirac
mass in this scenario. Using the basis $(\nu^c_R,\nu_R,\nu^{\prime
c}_R)$, which includes two SM singlet RH neutrinos with opposite
lepton numbers, the neutrino mass terms can be rewritten as
\begin{equation}
- {\cal L}_{m_\nu} = \frac{1}{2} \left(
\begin{array}{ccc} \bar{\nu}_L & \bar{\nu}^c_L &
\bar{\nu}^{\prime}_L \end{array} \right) \! \left(
\begin{array}{ccc}
0 & m_D & m^\prime \\ m_D & m_R & \mu^\prime_D \\
m^\prime & \mu^\prime_D & 0
\end{array} \right)
\left( \begin{array}{c} \nu^c_R \\ \nu_R \\ \nu^{\prime c}_R
\end{array} \right) + \; h.c. \, .
\label{massL2}
\end{equation}

The $B-L$ symmetry, here seen as a residual presymmetry, is broken
by the Majorana mass $m_R$ and the mixing $m^\prime$, so that they
should be much smaller than $m_D$ and $\mu^\prime_D$ in the 't
Hooft's meaning of naturalness.\cite{tHooft} The $\mu^\prime_D$,
coupling the sterile neutrinos, being independent of the vacuum
expectation value of the neutral component of the SM Higgs
doublet, and permitted by gauge invariance after the charge
normalization that converts preleptons into leptons, can take in
principle any value. Indeed, it has to be large because the
partnership of the ordinary RH neutrino in the presymmetric
arrangement is with the active neutrino and not with the extra
sterile neutrino, which appears as an outsider, a partner rejected
by presymmetry. Actually, the limit $m_R,m^\prime \ll m_D \ll
\mu^\prime_D$ corresponds to the linear seesaw
case,\cite{Linear1}$^{\mbox{--}}$\cite{Linear3} in which the mass
matrix can be block diagonalized with mass eigenvalues having the
order of magnitude given by
\begin{eqnarray}
\begin{array}{l}
\displaystyle m_1 \simeq - \frac{2m_D m^\prime}{\mu^\prime_D} +
\frac{m_R m^{\prime 2}}{\mu^{\prime 2}_D} , \\ [10pt] \displaystyle m_2
\simeq \mu^\prime_D + \frac{m_R}{2} +
\frac{(m_D+m^\prime)^2}{2\mu^\prime_D} +
\frac{m_R^2}{8\mu^\prime_D} , \\ [10pt] \displaystyle m_3
\simeq - \mu^\prime_D + \frac{m_R}{2} -
\frac{(m_D-m^\prime)^2}{2\mu^\prime_D} -
\frac{m_R^2}{8\mu^\prime_D} ,
\end{array}
\label{masses2}
\end{eqnarray}
where only the leading terms in $m^\prime$, $m_D$, $\mu^\prime_D$,
and $m_R$ are shown. These mass eigenvalues describe the light
mass of a Majorana active neutrino linear in $m_D$, explaining the
approach's name, and the heavy mass of a pseudo-Dirac pair with
mass splitting $\Delta m \simeq m_R + 2 m_D m^\prime /
\mu^\prime_D$, which can be so small as the mass of the active
neutrino. At the leading order, the active neutrino mass is
suppressed by its coupling $m^\prime$ to the extra sterile
neutrino, and in this case, also by the scale of the seesaw
mechanism. Note that the limit $m^\prime \rightarrow 0$, as in the
conventional inverse seesaw mechanism after spontaneous symmetry
breaking,\cite{Inverse1}$^{\mbox{--}}$\cite{Inverse3} leads to a massless
Majorana neutrino. Also observe that the mass matrix in
Eq.~(\ref{massL2}) cannot be rotated to the form given by the
inverse seesaw model because of presymmetry, which enforces the
extra neutrino Majorana mass $m^\prime_R = 0$; i.e. presymmetry
does not allow $\nu_R$ and $\nu^{\prime c}_R$ to be redefined by a
rotation so that $m^\prime=0$ and $m^\prime_R \neq 0$.\cite{Ma}
That is, from the perspective of presymmetry, the inverse seesaw
mechanism is not an option to have a low-scale seesaw, favoring
the linear variant instead.

Regarding the Majorana mass eigenstates
$\nu_{iM}=\nu_{iL}+\nu^c_{iR}$, there is an almost maximal mixing
between the RH neutrino and the extra sterile LH neutrino
according to
\begin{eqnarray}
\begin{array}{l}
\displaystyle \nu_{2L} \simeq \frac{1}{\sqrt{2}} \, (\nu^c_L+\nu^\prime_L) +
\frac{m_D}{\sqrt{2} \, \mu^\prime_D} \, \nu_L \, , \\ [10pt] \displaystyle
\nu_{3L} \simeq \frac{1}{\sqrt{2}} \, (\nu^c_L-\nu^\prime_L) -
\frac{m_D}{\sqrt{2} \, \mu^\prime_D} \, \nu_L \, ,
\end{array}
\label{Diracstates2}
\end{eqnarray}
and
\begin{eqnarray}
\begin{array}{l}
\displaystyle \nu^c_{2R} \simeq \frac{1}{\sqrt{2}} \, (\nu_R+\nu^{\prime
c}_R) + \frac{m_D}{\sqrt{2} \, \mu^\prime_D} \, \nu^c_R \, ,
\\ [10pt] \displaystyle \nu^c_{3R} \simeq \frac{1}{\sqrt{2}} \,
(\nu_R-\nu^{\prime c}_R) - \frac{m_D}{\sqrt{2} \, \mu^\prime_D} \,
\nu^c_R \, ,
\end{array}
\end{eqnarray}
which represent the Majorana components of a pseudo-Dirac neutrino
with heavy mass $\mu^\prime_D$ at the leading order. This
heavy pseudo-Dirac neutrino is mainly a pair made up of the
sterile neutrinos, so adulterating the ordinary pseudo-Dirac
neutrino by means of a mixture that has the sterile LH neutrino
instead of the active one as the principal constituent. Concerning
the light Majorana neutrino of mass $m_\nu \sim m_D m^\prime /
\mu^\prime_D$, we find
\begin{equation}
\nu_{1L} \simeq \nu_L - \frac{m^\prime}{\mu^\prime_D} \, \nu^c_L -
\frac{m_D}{\mu^\prime_D} \, \nu^\prime_L \, ,
\end{equation}
and
\begin{equation}
\nu^c_{1R} \simeq \nu^c_R - \frac{m^\prime}{\mu^\prime_D} \, \nu_R
- \frac{m_D}{\mu^\prime_D} \, \nu^{\prime c}_R \, ,
\label{Diracstate2s}
\end{equation}
where we have included the small mixing with sterile neutrinos. It
is seen that the mixing of order $m_D / \mu^\prime_D$ with the
extra sterile neutrino is much more significant than the mixing of
order $m^\prime / \mu^\prime_D \sim m_\nu / m_D$ with the RH
neutrino.

Well above the mass scale of $\nu_L$, we can integrate this out
using the equation of motion
\begin{equation}
\frac{d {\cal L}_{m_\nu}}{d \nu^c_R} = 0 , \label{motionEqlinear}
\end{equation}
which leads to
\begin{equation}
m_D \bar{\nu}^c_L + m^\prime \bar{\nu}^\prime_L = 0
\label{lineardecoupled}
\end{equation}
and the effective Lagrangian
\begin{equation}
- {\cal L}_{m_\nu} = \displaystyle \frac{1}{2} \left(
\begin{array}{cc} \bar{\nu}^c_L \; & \bar{\nu}^\prime_L
\end{array} \right) \left(
\begin{array}{cc}
m_{RR} \; &  m^{\prime}_{RL} \\
m^\prime_{RL} & m^{\prime}_{LL}
\end{array} \right)
\left( \begin{array}{c} \nu_R \\ \nu^{\prime c}_R
\end{array} \right) + \; h.c.,
\label{Lefflinear}
\end{equation}
where
\begin{equation}
\displaystyle m_{RR} = m_R + \frac{m_D m^\prime}{\mu^\prime_D} ,
\qquad m^{\prime}_{LL} = \frac{m_D m^\prime}{\mu^\prime_D} ,
\qquad \displaystyle m^{\prime}_{RL} = \mu^\prime_D + \frac{m^2_D
+ m^{\prime 2}}{2 \mu^\prime_D}. \label{linearmatrix}
\end{equation}
It is worth mentioning that to completing Eq.~(\ref{Lefflinear}),
the following expressions obtained from
Eq.~(\ref{lineardecoupled}) need to be added:
\begin{eqnarray}
\begin{array}{l}
\displaystyle \frac{m_D m^\prime}{2 \mu^\prime_D} \bar{\nu}^c_L \nu_R +
\frac{m^{\prime 2}}{2 \mu^\prime_D} \bar{\nu}^\prime_L \nu_R +
h.c. = 0 , \\ [10pt] \displaystyle \frac{m^2_D}{2 \mu^\prime_D}
\bar{\nu}^\prime_L \nu_R + \frac{m_D m^\prime}{2 \mu^\prime_D}
\bar{\nu}^\prime_L \nu^{\prime c}_R + h.c. = 0 .
\end{array}
\end{eqnarray}

The mass eigenvalues of the mass matrix in Eq.~(\ref{Lefflinear})
are in agreement with the heavy masses of the pseudo-Dirac pair
displayed in Eq.~(\ref{masses2}), which have a mass splitting
given by $\Delta m = m_{RR} + m^\prime_{LL}$. We note
that Eqs.~(\ref{Lefflinear}) and (\ref{linearmatrix}) are also
gotten through the block diagonalization of the mass matrix in
Eq.~(\ref{massL2}).

On the other hand, if we assume a bottom-up framework where the
residual presymmetry $U(1)_{B-L}$ is violated only in couplings of
the extra sterile neutrino, implying $m_R=0$, Eq.~(\ref{masses2})
is simplified to
\begin{eqnarray}
\begin{array}{l}
\displaystyle m_1 \simeq - \frac{2m_D m^\prime}{\mu^\prime_D} ,
\\ [10pt] \displaystyle m_2 \simeq \mu^\prime_D +
\frac{(m_D+m^\prime)^2}{2\mu^\prime_D} , \\ [10pt]
\displaystyle m_3 \simeq - \mu^\prime_D -
\frac{(m_D-m^\prime)^2}{2\mu^\prime_D} ,
\end{array}
\label{masses3}
\end{eqnarray}
as in the usual linear seesaw after spontaneous symmetry
breaking,\cite{Linear1}$^{\mbox{--}}$\cite{Linear3} with the heavy
pseudo-Dirac pair having a significant mass splitting, related to
the small mass of the active neutrino. The Dirac limit
$m^\prime = m_R = 0$ produces a massless active neutrino and a heavy
Dirac fermion with the sterile neutrinos, with mass of order
$\mu^\prime_D + m^2_D / 2\mu^\prime_D$. It restores
$B-L$ symmetry, which is conceived of a residual presymmetry.

Benchmarks on the model are set down in Sec.~\ref{constraints},
where masses are constrained from experimental limits on the mass
splitting in pseudo-Dirac neutrinos, pursuing a method not
employed before.
%%%%%%%%%%%%%%%%%%%%%%%%%%%%%%%%%%%%%%%%%%%%%%
\section{Presymmetry in High-Scale Seesaw with Light \\
Pseudo-Dirac Neutrinos} \label{normalseesaw}

Here we consider the inverted alternative to the model of the
heavy pseudo-Dirac pair seen in Sec.~\ref{linearseesaw}, which is
constructed with the extra gauge singlet fermion that implements
the low-scale seesaw mechanism. Two RH neutrinos per generation,
having the same standard quantum numbers at the lepton--quark
level, are now included in order to adulterate the neutrino mass
by suppressing the ordinary one ($\nu_R$) through the high-scale
seesaw and generating a pseudo-Dirac neutrino of tiny mass with
the other ($\nu^\prime_R$). The extra RH neutrino is a gauge
singlet with trivial topology and much weaker couplings. The
naturalness in the 't Hooft's sense of this smallness is related
to the presymmetry between leptons and quarks with one RH
neutrino.

The realization of light neutrinos as a pseudo-Dirac pair of an
active LH and an extra sterile RH neutrino was described in
Refs.~\refcite{pDirac1} and \refcite{pDirac2}. Even though we are
aware that this mass spectrum is in tension with cosmological
bounds, we revise this model to show how the new implementation of
presymmetry leads to the light neutrinos. The procedure we follow
has a close resemblance to that described in
Sec.~\ref{linearseesaw}. The new Yukawa Lagrangian to start with
in the context of presymmetry is
\begin{equation}
- {\cal L}^{\prime}_{\nu} = y \, \bar{\ell}_L \tilde{\phi} \nu_R +
y^\prime \, \bar{\ell}_L \tilde{\phi} \nu^\prime_R + \frac{1}{2}
\left( \begin{array}{cc} \bar{\nu}^c_L & \bar{\nu}^{\prime c}_L
\end{array} \right) \left(
\begin{array}{cc}
m_R & \mu^\prime \\
\mu^\prime & 0
\end{array} \right)
\left( \begin{array}{c} \nu_R \\ \nu^\prime_R
\end{array} \right)
+ h.c. \, , \label{YukawaL}
\end{equation}
where the Majorana mass terms with $m_R$ and $\mu^\prime$ are
included. They describe transitions between RH neutrinos and
conjugate LH antineutrinos, reflecting soft symmetry breaking of
$U(1)_{B-L}$ and can be generated by the vacuum expectation value
of a gauge singlet scalar coupled to the RH neutrinos.

The Yukawa coupling $y^\prime$ of the extra RH neutrino has to be
small in comparison with the $y$ of the ordinary RH neutrino
because the partner of the active neutrino in the presymmetric
arrangement is the ordinary one and not the extra, which becomes
an outsider, a partner other than the one recognized by
presymmetry. That is, the presymmetric Lagrangian that underlies
Eq.~(\ref{YukawaL}) is just that written in
Eq.~(\ref{preyukawa2}); since the extra RH neutrino is free
from any kind of gauge charge, the Yukawa term $\bar{\hat{\ell}}_L
\tilde{\phi} \nu^\prime_R$ is not admitted by gauge invariance. We
remark that its inclusion would break the fractional charge
presymmetry between the lepton and quark sectors. On the other
hand, the couplings $\mu^\prime$ and $m_R$, involving the ordinary
RH but not the LH neutrino and allowed by gauge invariance after
the charge normalization that transforms preleptons into leptons,
can have in principle any value. Here we assume a high-scale
seesaw, i.e. $\mu^\prime \ll m_R$, having so a large difference
between the Majorana mass of RH neutrinos.

After the spontaneous breaking of the electroweak gauge group of
the SM, the Dirac neutrino masses are induced through
$m_D=y\langle \phi^0 \rangle$ and $m^\prime_D=y^\prime \langle
\phi^0 \rangle$. Using the basis $(\nu^c_R,\nu_R,\nu^\prime_R)$,
the neutrino mass terms with $m^\prime_R =0$ can be written as
\begin{equation}
- {\cal L}_{m_\nu} = \frac{1}{2} \left(
\begin{array}{ccc} \bar{\nu}_L & \bar{\nu}^c_L & \bar{\nu}^{\prime
c}_L \end{array} \right) \left(
\begin{array}{ccc}
0 & m_D & m^\prime_D \\ m_D & m_R & \mu^\prime \\
m^\prime_D & \mu^\prime & 0
\end{array} \right)
\left( \begin{array}{c} \nu^c_R \\ \nu_R \\ \nu^\prime_R
\end{array} \right) + h.c. \, .
\label{massL}
\end{equation}

The mass matrix can be block diagonalized through a seesaw
mechanism. Assuming the parameter space $m_D, m^\prime_D,
\mu^\prime \ll m_R$ as in a normal seesaw, and $m^2_D/m_R,
\mu^{\prime 2}/m_R \ll m^\prime_D$ for a pseudo-Dirac regime, the
mass eigenvalues are
\begin{eqnarray}
\begin{array}{l}
\displaystyle m_1 \simeq m^\prime_D - \frac{1}{2}
\frac{(m_D+\mu^\prime)^2}{m_R} , \\ [10pt] \displaystyle m_2
\simeq - m^\prime_D - \frac{1}{2}
\frac{(m_D-\mu^\prime)^2}{m_R} , \\ [10pt]
\displaystyle m_3 \simeq m_R + \frac{m^2_D+\mu^{\prime 2}}{m_R} ,
\end{array}
\label{masses}
\end{eqnarray}
where only the leading terms in $m^\prime_D$, $m_D$, $\mu^\prime$,
and $m_R$ are indicated. These expressions show that the large
$m_R$ gives the mass of a heavy Majorana neutrino and fixes the
mass splitting in a light pseudo-Dirac pair. The parameters $m_D$
and $\mu^\prime$ are suppressed by $m_R$, but $m^\prime_D$ is not.
Thus, the small mass of light neutrinos fixed by $m^\prime_D$ does
not depend on the high-scale of the seesaw mechanism.

The three Majorana mass eigenstates are
$\nu_{iM}=\nu_{iL}+\nu^c_{iR}$ ($i=1,2,3$), where
\begin{eqnarray}
&& \nu_{1L} \simeq \frac{1}{\sqrt{2}}(\nu_L+\nu^{\prime c}_L) -
\frac{m_D}{\sqrt{2} \, m_R} \, \nu^c_L \, , \nonumber \\
&& \nu_{2L} \simeq \frac{1}{\sqrt{2}}(-\nu_L+\nu^{\prime c}_L) +
\frac{m_D}{\sqrt{2} \, m_R} \, \nu^c_L \, , \\
&& \nu_{3L} \simeq \nu^c_L + \frac{m_D}{m_R} \, \nu_L +
\frac{\mu^\prime}{m_R} \, \nu^{\prime c}_L \, , \nonumber
\label{Diracstates}
\end{eqnarray}
and
\begin{eqnarray}
&& \nu^c_{1R} \simeq \frac{1}{\sqrt{2}}(\nu^c_R+\nu^\prime_R) -
\frac{m_D}{\sqrt{2} \, m_R} \, \nu_R \, , \nonumber \\
&& \nu^c_{2R} \simeq \frac{1}{\sqrt{2}}(-\nu^c_R+\nu^\prime_R) +
\frac{m_D}{\sqrt{2} \, m_R} \, \nu_R \, , \\
&& \nu^c_{3R} \simeq \nu_R + \frac{m_D}{m_R} \, \nu^c_R +
\frac{\mu^\prime}{m_R} \, \nu^\prime_R \, , \nonumber
\label{conjugatestates}
\end{eqnarray}
which are analogous to those given in
Eqs.~(\ref{Diracstates2})--(\ref{Diracstate2s}). It is then shown
that there is an almost maximal mixing of the active LH neutrino
$\nu_L$ with the extra RH neutrino $\nu^\prime_R$, and a strong
suppression of its mixing with its natural partner $\nu_R$, i.e.
the light pseudo-Dirac neutrino is mainly a pair made up of the
active neutrino and the extra sterile neutrino, so adulterating
the ordinary pseudo-Dirac neutrino generated with the ordinary RH
neutrino.

The field $\nu_R$ approximately becomes a heavy mass eigenstate
and is decoupled at low energies. It can be integrated out solving
the equation of motion
\begin{equation}
\frac{d {\cal L}_{m_\nu}}{d \nu_R} = 0 , \label{motionEq}
\end{equation}
which leads to
\begin{eqnarray}
\begin{array}{l}
\displaystyle \bar{\nu}^c_L = - \frac{m_D}{m_R} \bar{\nu}_L -
\frac{\mu^\prime}{m_R} \bar{\nu}^{\prime c}_L , \\ [10pt] \displaystyle
\nu_R = - \frac{m_D}{m_R} \nu^c_R - \frac{\mu^\prime}{m_R}
\nu^\prime_R ,
\end{array}
\label{decoupled}
\end{eqnarray}
and the effective Lagrangian
\begin{equation}
- {\cal L}_{m_\nu} = \displaystyle \frac{1}{2} \left(
\begin{array}{cc} \bar{\nu}_L \; & \bar{\nu}^{\prime c}_L
\end{array} \right) \left(
\begin{array}{cc}
m_{LL} \; &  m^{\prime}_{LR} \\
m^\prime_{LR} & m^{\prime}_{RR}
\end{array} \right)
\left( \begin{array}{c} \nu^c_R \\ \nu^\prime_R
\end{array} \right) + h.c.,
\label{Leff}
\end{equation}
where
\begin{equation}
\displaystyle m_{LL} = -\frac{m^2_D}{m_R} , \qquad m^{\prime}_{RR}
= -\frac{\mu^{\prime 2}}{m_R} , \qquad \displaystyle
m^{\prime}_{LR} = m^\prime_D - \frac{m_D\mu^\prime}{m_R} .
\label{matrix}
\end{equation}

The mass eigenvalues correspond to the light masses of the
pseudo-Dirac pair given in Eq.~(\ref{masses}) with mass splitting
$\Delta m \simeq (m^2_D+\mu^{\prime2})/m_R$ depending on the
diagonal terms $m_{LL}$ and $m^\prime_{RR}$, the effective
Majorana mass parameters. Here we again note that this effective
Lagrangian is also obtained by means of the block diagonalization
of the mass matrix in Eq.~(\ref{massL}).
Equations~(\ref{motionEq})--(\ref{matrix}) for the light
pseudo-Dirac neutrino are analogous to
Eqs.~(\ref{motionEqlinear})--(\ref{linearmatrix}) for the heavy
pseudo-Dirac neutrino.

Within the pseudo-Dirac framework with one RH neutrino, a Dirac
neutrino mass $m_D$ of small value compared with charged leptons
is considered unnatural. In our extended scenario, however, a
small Dirac neutrino mass $m^\prime_D$ becomes natural because
presymmetry between leptons and quarks is primarily established
with the ordinary RH neutrino. Therefore pseudo-Dirac neutrinos
with small masses can be accommodated naturally.

The Dirac limit $\Delta m \simeq (m^2_D + \mu^{\prime 2})/m_R
\rightarrow 0$ is natural as the symmetry enhances by the global
$B-L$ symmetry, which is viewed as a residual presymmetry. In this
case, Eq.~(\ref{Leff}) becomes
\begin{equation}
- {\cal L}_{m^\prime_\nu} = m^\prime_D (\bar{\nu}_L \nu^\prime_R +
\bar{\nu}^\prime_R \nu_L) \, , \label{DiracL}
\end{equation}
which represents a light Dirac neutrino with the
small mass $m^\prime_D \ll m_D$ and with the extra RH neutrino
as its sterile component. The explanation for the strong
hierarchy between the Dirac masses is provided by presymmetry with
one RH neutrino, which also guarantees the stability of the small
mass under corrections from the quantum field theory and the
interplay with the seesaw mechanism.

Finally, it is worth noting that the global $\mbox{U}(1)_{B-L}$
symmetry used in this rather conventional presentation of
presymmetry is not an accidental symmetry. In contrast, it is a
distinctive feature of presymmetry which forbids the Majorana mass
terms of extra sterile neutrinos added to the particle content.

Experimental constraints on the mass parameters of the model are
given below, where we depart from our usual procedure of bounding
them.
%%%%%%%%%%%%%%%%%%%%%%%%%%%%%%%%%%%%%%%%%%%%%%
\section{On Constraints Facing Pseudo-Dirac Neutrinos in \\
Seesaw with Low/High Scale} \label{constraints}

In the low-scale linear seesaw with heavy pseudo-Dirac and light
Majorana neutrinos, the extension of the SM is done with one RH
and one LH sterile neutrinos per generation. The choice is to use
the seesaw mechanism to decoupling these two sterile neutrinos
from the other by making the coupling between them much higher
than the other mass parameters.

For simplicity, we work in the approximation of one generation
assuming $m_R=0$. In order to constrain masses we rewrite the mass
eigenvalues in Eq.~(\ref{masses3}) as
\begin{eqnarray}
&& m_1 \simeq - \Delta m , \nonumber \\
&& m_2 \simeq \mu^\prime_D + \frac{m^2_D+m^{\prime
2}}{2\mu^\prime_D} + \frac{\Delta m}{2} , \\
&& m_3 \simeq - \mu^\prime_D - \frac{m^2_D+m^{\prime
2}}{2\mu^\prime_D} + \frac{\Delta m}{2} , \nonumber
\end{eqnarray}
while from Eq.~(\ref{linearmatrix}) we have
\begin{equation}
m_{RR} \simeq m^\prime_{LL} \simeq \frac{\Delta m}{2} , \qquad
m^\prime_{RL} \simeq \mu^\prime_D + \frac{m^2_D+m^{\prime
2}}{2\mu^\prime_D} ,
\end{equation}
where $\Delta m \simeq 2 m_D m^\prime / \mu^\prime_D$ is the
mass splitting in the heavy pseudo-Dirac neutrino. Thus, $\Delta
m = |m_2|-|m_3|\simeq |m_1|$ decreases and $|m_{2,3}|$ increase
with increasing $\mu^\prime_D$. We therefore have a seesaw
mechanism essentially involving the mass splitting in the heavy
pseudo-Dirac neutrino rather than the mass $m_1$, which decreases
with increasing $\mu^\prime_D$. The Dirac limit
$\Delta m \rightarrow 0$ with conservation of $B-L$ generates a
massless active neutrino and a heavy Dirac neutrino. The hierarchy
$m_\nu \ll m_D$ is then natural in the 't~Hooft's sense.

Using the condition $m^\prime \ll m_D$ and the experimental data
on the mass splitting $\Delta m \simeq m_{RR} + m^\prime_{LL}
\simeq |m_1|$,\cite{PDG} namely
\begin{equation}
\Delta m \simeq m_\nu \simeq 10^{-1} \mbox{eV} ,
\end{equation}
we have a low-energy threshold for the seesaw given by
\begin{equation}
\mu^\prime_D \ll \frac{2 m^2_D}{m_\nu} \simeq 2 \times 10^{10}
(\frac{m^2_D}{\mbox{1 GeV}}) ,
\label{constraint2}
\end{equation}
which depends on the value chosen for $m_D$. Assuming the Dirac
mass of order the charged lepton mass, $m_D \simeq \mbox{1 MeV}$
for the first generation. This yields $\mu^\prime_D \ll 20 \;
\mbox{TeV}$, allowing heavy pseudo-Dirac neutrinos in the TeV
range. For illustrative purposes, if $\mu^\prime_D \simeq 1 \;
\mbox{TeV}$, we get
\begin{equation}
m^\prime \simeq \frac{m_\nu \mu^\prime_D}{2 \, m_D} \simeq 5
\times 10^{-2} \; \mbox{MeV},
\end{equation}
and the neutrino masses
\begin{equation}
|m_1| \simeq 10^{-1} \; \mbox{eV} , \qquad |m_2| \simeq |m_3|
\simeq 1 \; \mbox{TeV} .
\end{equation}
In the case of $m_D \simeq 100 \, \mbox{GeV}$, the seesaw scale is
now $\mu^\prime_D \ll 2 \times 10^{14} \; \mbox{GeV}$, a value
below the scale of grand unified theories.

The seesaw with light Majorana and heavy pseudo-Dirac neutrinos
has received a great deal of attention because it can be tested at
low energies, in collider and non-collider experiments, such as
neutrinoless double beta decay, lepton flavor violation processes,
nonunitarity, neutrino experiments, rare-meson decays, etc. For a
review see for instance Ref.~\refcite{Boucenna}. For recent work
see e.g. Refs.~\refcite{recent1}--\refcite{recent3} and references
therein. A successful leptogenesis to solve the problem of the
baryon asymmetry in the universe is presented in
Ref.~\refcite{leptogenesis}.

On the other hand, in the high-scale seesaw with light
pseudo-Dirac and heavy Majorana neutrinos, the inverted
alternative where the extension of the SM is done with two RH
neutrinos in each generation, the first choice is to use the
seesaw mechanism to decoupling the original RH neutrinos from the
others by making them much heavier than the other mass parameters
and ratios of mass parameters. The second one is to use the Dirac
mass of the extra RH neutrinos to control effectively the light
neutrino masses. This demands a hierarchy between the extra Dirac
masses and the other mass factors. This second hierarchy brings
out the light pseudo-Dirac neutrinos.

Working in the approximation of one generation, the mass
eigenvalues in Eq.~(\ref{masses}) come to be
\begin{eqnarray}
\begin{array}{l}
\displaystyle m_1 \simeq m^\prime_D - \frac{m_D\mu^\prime}{m_R} -
\frac{\Delta m}{2} , \\ [10pt] \displaystyle
m_2 \simeq - m^\prime_D + \frac{m_D\mu^\prime}{m_R} -
\frac{\Delta m}{2} , \\ [10pt] \displaystyle
m_3 \simeq m_R + \Delta m ,
\end{array}
\end{eqnarray}
whereas the mass matrix elements in Eq.~(\ref{matrix}) become
\begin{equation}
m_{LL} + m^\prime_{RR} \simeq - \Delta m , \qquad  m^\prime_{LR}
\simeq m^\prime_D - \frac{m_D\mu^\prime}{m_R} ,
\end{equation}
with $\Delta m \simeq (m^2_D+\mu^{\prime 2})/m_R$ being now the
mass splitting in the light pseudo-Dirac neutrino. Hence, $\Delta
m = |m_2|-|m_1|$ decreases and $m_3$ increases with increasing
$m_R$. We also have here a seesaw mechanism essentially involving
the mass splitting in the light pseudo-Dirac neutrino rather than
its $m_{1,2}$ masses, which increase with increasing $m_R$. Being
the heavy Majorana neutrino decoupled, the Dirac limit $\Delta m
\rightarrow 0$ ($m_R \rightarrow \infty$) with conservation of
$B-L$ is clearly exhibited. The hierarchy $\Delta m \ll
m^\prime_D$ is then natural in the 't~Hooft's meaning. Note that,
using our parameter regime, $m_D\mu^\prime/m_R$ is also much
smaller than $m^\prime_D$ in $m_{1,2}$ and $m^\prime_{LR}$.

Current data on solar neutrino oscillations yield an upper bound
on the mass splitting $\Delta m \simeq
|m_{LL}+m^\prime_{RR}|$,\cite{Gouvea}
\begin{equation}
\Delta m < 10^{-9} \, \mbox{eV} ,
\end{equation}
which, taking $\mu^\prime < m_D$, sets the lower limit
\begin{equation}
m_R > 10^{18} (\frac{m^2_D}{\mbox{1 GeV}}) . \label{constraint1}
\end{equation}
This is much higher than the threshold of $\mu^\prime_D$ given in
Eq.~(\ref{constraint2}). For $m_D \simeq \mbox{1 MeV}$, it leads
to $m_R > 10^{12} \; \mbox{GeV}$ which is not greatly different
from the scale of grand unified theories. The benchmarks on
neutrino masses are then
\begin{equation}
|m_1| \simeq |m_2| \simeq m^\prime_D \simeq m_\nu \simeq 10^{-1}
\; \mbox{eV} , \qquad |m_3| \simeq m_R > 10^{12} \; \mbox{GeV} ,
\end{equation}
where $m_\nu$ refers to the laboratory data on neutrino mass. We
note that if $m_D \simeq 100 \, \mbox{GeV}$, i.e. at the
electroweak scale, a threshold greater than the Planck scale is
obtained. Although extremely small, a $\Delta m$ different from
zero would imply that the light neutrino is a small perturbation
of the Dirac case, consistent with an almost exact $B-L$ symmetry,
which is regarded as a residual presymmetry.

There is a number of works provided by different groups that have
studied the high-scale seesaw with heavy Majorana and light
pseudo-Dirac neutrinos. The tiny neutrino masses as well as the
baryon asymmetry of the universe, via leptogenesis, can be
explained. However, the very heavy mass of Majorana neutrinos and
the very tiny mass splitting in pseudo-Dirac neutrinos make them
hard to probe in low energy experiments. There exists a decoupling
at low energies of the effects from the heavy Majorana neutrinos,
other than providing mass to light neutrinos, and a direct
signature at colliders becomes impossible as well as the
observation of induced lepton flavor violation processes.
Regarding the light pseudo-Dirac neutrinos, only future neutrino
telescopes could offer a way to detect their effects by observing
the long-wavelength oscillations between nearly degenerate active
neutrinos and sterile ones with the same chirality during their
travel from distant astrophysical sources like active galactic and
extra-galactic nuclei. See e.g. Ref.~\refcite{NeTelescope} and
references therein.
%%%%%%%%%%%%%%%%%%%%%%%%%%%%%%%%%%%%%%%%%%%%%%
\section{Conclusions}
\label{conclusions}

We have implemented presymmetry in a rather conventional form by
imposing the global $U(1)_{B-L}$ symmetry on the theory of leptons
and quarks with one RH neutrino per generation and symmetric bare
electroweak charges. The $B-L$ symmetry favors the case of
symmetric fractional quantum numbers as in quarks, so that
fractional charges are ``hidden'' not only in hadrons but also in
leptons. However, the fractional charges of the preleptons
postulated by the presymmetry model do not correspond to physical
charges measured in the laboratory. They are bare charges that do
not take into account the contributions from the topological
charges generated by counterterms that cancel out the troublesome
gauge anomalies produced by the fractional charges. The
cancellation of these divergences and the non-perturbative
normalization of prelepton charges proceed via the mixing of local
and topological charges, the global $B-L$ measuring the induced
charge from a unit of topological charge, and the mathematical
replacement of the initially postulated fractional charges with
the experimentally observed charges. Thus, the charge splitting in
the lepton sector is unphysical and does not have observable
effects.

Models of massive neutrinos proposed in the literature were
examined from the perspective of presymmetry. Thus, the
$U(1)_{B-L}$ symmetry of the SM extended with RH neutrinos and
Yukawa couplings as those of charged leptons is viewed as a
residual presymmetry. No residual effects of presymmetry, however,
remain in high-scale seesaw models with Majorana mass terms for
the RH neutrinos. To have a low-scale seesaw, a second set of
sterile neutrinos is added to the particle content with almost
maximal mixing with the other neutrinos. The distinguishing
features of the presymmetry approach are now the forbiddance of
the Majorana mass of the extra sterile neutrinos and their nonzero
couplings to the active neutrinos. The small values of these
couplings fix the mass scale of light neutrinos and the mixing of
the active neutrinos with the extra sterile neutrinos is much more
significant than their mixing with the RH neutrinos.

The low-scale linear seesaw mechanism leading to light Majorana
neutrinos and heavier pseudo-Dirac neutrinos, and its inverted
high-scale alternative generating light pseudo-Dirac neutrinos and
heavy Majorana neutrinos, were considered, where these
pseudo-Dirac particles receive their masses dominantly from Dirac
mass terms involving the adulterant, extra sterile neutrinos with
the imprints of presymmetry. We have found that in these seesaws,
the tiny mass splitting in pseudo-Dirac neutrinos and the mass of
heavy neutrinos ride on the opposite ends, so that
phenomenological bounds on the splitting necessarily imply
adjustments of the seesaw scale by means of the Majorana and Dirac
mass terms. Although many experiments to probe the Majorana or
Dirac character of light and heavy neutrinos have been realized,
no conclusive evidence for them has been found yet, so that a
pseudo-Dirac property is still a possibility. The pseudo-Dirac
character of light neutrinos, however, would be hard to probe if
the splitting is very tiny, without considering the tension that
this mass spectrum has with cosmological bound on the number of
relativistic neutrino degrees of freedom.

We have shown that the presymmetry scenario discards the
conventional inverse seesaw, which assumes no couplings between
the extra sterile neutrinos and the active neutrinos,
favoring the linear variant instead. Presymmetry does not allow the
fermionic singlets to be redefined by a rotation to relate the
linear and inverse seesaw models, as usually done.

The relevance of presymmetry and sterile neutrinos in the
mechanisms of neutrino mass generation are readily perceived,
independently whether light neutrinos are Dirac, Majorana or
pseudo-Dirac fermions, questions that are currently under intense
experimental investigations. Yet, the seesaw framework used here
for presymmetry is not enough to explaining the pattern of the
observed neutrino mixing matrix. A way to treat this issue is to
extend the SM with a discrete flavor symmetry, which may be
supplemented with an extension of the gauge symmetry, including a
gauged presymmetry, so promoting the global $U(1)_{B-L}$ of
presymmetry to a local symmetry. The study of a specific model,
however, is beyond the scope of this paper.
%%%%%%%%%%%%%%%%%%%%%%%%%%%%%%%%%%%%%%%%%%%%%%
\section*{Acknowledgments}

This work was supported by Vicerrector\'{\i}a de Investigaci\'on,
Desarrollo e Innovaci\'on, Universidad de Santiago de Chile,
Usach, Proyecto DICYT No. 041731MC.
%%%%%%%%%%%%%%%%%%%%%%%%%%%%%%%%%%%%%%%%%%%%%%

\end{document}